# Is the Free Vacuum Energy Infinite?

**H. Razmi** [(1)] and **S. M. Shirazi** [(2)]

Department of Physics, the University of Qom, Qom, I. R. Iran.
(1) razmi@qom.ac.ir & razmiha@hotmail.com (2) sms102@gmail.com

## Abstract

Considering the fundamental cutoff applied by the uncertainty relations' limit on virtual particles' frequency in the quantum vacuum, it is shown that the vacuum energy density is proportional to the inverse of the forth power of the dimensional distance of the space under consideration and thus the corresponding vacuum energy automatically regularized to zero value for an infinitely large free space. This can be used in regularizing a number of unwanted infinities happen in the Casimir effect, the cosmological constant problem and so on without using already known mathematical (not so reasonable) techniques and tricks.

**Keywords:** Quantum Vacuum; the Uncertainty relations; Virtual Particles

**PACS:** 03.70.+k; 42.50.Pq

## Introduction

In the standard quantum field theory, not only the vacuum (zero point) energy has an absolute infinite value, but also all the real excited states have such an irregular value; this is because these energies corresponds to the zero-point energy of an infinite number of harmonic oscillators ($W = \frac{1}{2}\sum_{k,\sigma}\hbar\omega_k \to \infty$). We usually get rid of this irregularity via simple technique of normal ordering by considering the energy difference relative to the vacuum state [1-6]; but of course, there are some important situations where one deals directly with the absolute vacuum energy as in the cosmological constant problem [7], or in the regularization of the vacuum energy in the Casimir effect [8]. After a half century of knowing and "living" with the vacuum energy, there are only some mathematical techniques and approaches in regularizing its infinite value without paying enough conceptual attention to the "content" of the quantum vacuum "structure". In this paper, considering the fundamental assumption that the vacuum energy originates from the "motion" of virtual particles in the quantum vacuum, it is shown that the free vacuum energy can be regularized based on the uncertainty relations' limit on these particles' frequency. Indeed, the free vacuum (or any infinitely large vacuum space) energy is automatically regularized to zero value without using any presupposition (e.g. normal ordering) usually used in getting rid of the infinity of the quantum vacuum energy (the vacuum catastrophe).

## The quantum vacuum, virtual particles and the uncertainty relations

The quantum vacuum is not really empty. It is filled with virtual particles which are in a continuous state of fluctuation. Virtual particle-antiparticle pairs are created from vacuum and annihilated back to it. These virtual particles exist for a time dictated by Heisenberg uncertainty relation. Based on the uncertainty relations, for any virtual particle, there is a limit on the timescale of "being" created from the vacuum fluctuations and then annihilated back to vacuum (its "lifetime"); thus, there should be a limit on the frequency of the virtual particles whose total energies is considered as the vacuum energy. In quantum (field) theory, it is well-known that the reason for naming the quantum vacuum particles as virtual particles is that although they are in "existence" and can have observable effects (e.g. the Casimir effect, spontaneous emission, Lamb shift), they cannot be directly detected (i.e. they are unobservable). For these unobservable (virtual) particles, the energy and lifetime values are constrained due to the uncertainty relation and can take, at most, the minimum values of uncertainties for real particles. This can be written as the following relation:

$$(E\tau)\big|_{\max.}^{virtual} \approx \alpha\frac{\hbar}{2} \qquad (1),$$

where $\alpha$ is a constant which cannot have a value much more than 1 to guarantee that we are dealing with virtual particles than real ones. As we know, the uncertainties in energy and lifetime of real (detectable) particles satisfy the relation:

$$\Delta E\Delta\tau \geq \frac{\hbar}{2} \qquad (2).$$

## The vacuum energy density of infinitely large free spaces

Although attribution of physical parameters and quantities to the virtual particles as the same as what we know for the real particles isn't a completely known and proved fact, the main reason of irregularity/infinity of the vacuum energy in QFT is because of attribution of the frequency $\omega$ to the virtual particles and summing on the infinite modes for them. Also, attributing "distance" to virtual particles is a known fact; in the Casimir effect, we say about the confinement of virtual particles in a finite distance between the two plates and the Casimir force depends on this distance. The only known point about the "attendance" of the virtual photons in a finite distance in QFT is that these intermediate particles (as in the Feynman diagrams) have nonzero masses with finite range of "action"; this makes them have a velocity of $v < c$ where we shall consider it in our calculation.

For a free space of dimensional length $D$, using the relations $E = \hbar\omega$, $\tau \leq \frac{D}{c}$, and (1), the frequency of virtual particles should satisfy:

$$\omega\Big|^{virtual} \leq \frac{\alpha c}{2D} \qquad (3).$$

Considering this limit on $\omega$ and the following well-known relation for the vacuum energy density corresponding to an infinitely large space:

$$\frac{E}{V} = \frac{1}{2V}\sum_k \hbar\omega_k \rightarrow 2\pi\hbar c\int k^3 dk \qquad (4),$$

it is found:

$$\frac{E}{V} = 2\pi\hbar c\int_0^{\frac{\alpha}{2D}} k^3 dk = \frac{\pi\hbar c\alpha^4}{8D^4} \quad (5).$$

For an infinitely large free space, the vacuum energy is zero; it is automatically regularized as in the following:

$$E_{free} = (\frac{\pi\hbar c\alpha^4}{8D^4})(Volume) \approx (\frac{\pi\hbar c\alpha^4}{8D^4})(D^3) = \frac{\pi\hbar c\alpha^4}{8D} \rightarrow 0 \qquad (D \rightarrow \infty) \qquad (6).$$

## Discussion

This result that the vacuum energy of the free infinitely large spaces is zero may be interpreted as that the infinite vacuum is a potentially resource containing infinitely free virtual particles of negligible frequency where can take higher values of frequency (energy) under the influence of the restrictions made on them by the presence of external boundaries that constrain their infinite freedom; this interpretation seems to be more reasonable than that the vacuum energy for the free infinitely large (or even finite) space has an infinite (irregular) value.

We should mention that the different meaning (interpretation) of energy-time uncertainty relation from the well-known momentum-position uncertainty principle doesn't affect what we have calculated here. Indeed, in the relation (2), $\Delta\tau$ means as a "lifetime" width quantity than as an uncertainty in time. It is also mentioned that the result of this paper isn’t in conflict to the response of quantum vacuum to a finite bounded restriction (the Casimir effect). Indeed, although the main “sound” of this paper is that the Casimir energy for free spaces or infinitely large outer spaces in the standard geometries well-known in the Casmir effect becomes zero in spite of already accepted infinite (irregular) values, it is possible to find out the expected Casimir force for the well-known problem of two parallel conducting plates based on the regularization introduced here (see **Appendix**).

## Appendix

For two plates of distance $d$ from each other, there is a freedom of $x \sim d$ for inner virtual particles and $x \sim D$ ($D \to \infty$) for the particles in the two (left and right) outer spaces. As is well-known, the Casimir energy corresponding to the famous geometry of two parallel conducting plates is:

$$E_{Casimir} = E_{bounded} - E_{free} = (E_{left} + E_{inside} + E_{right}) - E_{free} \qquad \text{(A-1).}$$

Considering the resulting relation (6), all three terms $E_{left}$, $E_{right}$, and $E_{free}$ vanish and thus:

$$E_{Casimir} = E_{inside} \qquad \text{(A-2).}$$

For the case of a scalar field [9], using the well-known energy-momentum tensor field

$$T^{\alpha\beta} = \partial^{\alpha}\varphi\partial^{\beta}\varphi - \frac{1}{2} g^{\alpha\beta}\partial_{\mu}\varphi\partial^{\mu}\varphi \qquad \text{(A-3),}$$

and the following relation between vacuum to vacuum expectation value of the field operators at two space-time points and the time dependent Green function (the propagator)

$$\langle 0 | T\{\varphi(x)\varphi(x')\} | 0 \rangle = -i\hbar c G(x,x') \qquad \text{(A-4),}$$

and the relation

$$\langle 0 | \hat{T}^{\alpha\beta} | 0 \rangle = -i\hbar c \lim_{x \to x'} (\partial^{\alpha}\partial'^{\beta} - \frac{1}{2} g^{\alpha\beta}\partial_{\mu}\partial'^{\mu}) G(x,x') \qquad \text{(A-5),}$$

by means of

$$G(x,x')_{in} = \frac{-1}{(2\pi)^3} \int \frac{d\omega}{c} d^2k e^{-i\omega(t-t')} e^{ik_{\perp}.(x-x')} \frac{1}{\lambda \sin \lambda d} \sin(\lambda z_{<}) \sin \lambda (z_{>} - d)$$

$$\lambda^2 = \frac{\omega^2}{c^2} - k^2, \vec{k}_{\perp} = \vec{k}_x + \vec{k}_y \equiv \vec{k} \qquad \text{(A-6),}$$

we arrive at this result that:

$$\left\langle T^{00}\right\rangle_{in} = \frac{i\hbar c}{2(2\pi)^3}\int \frac{d\omega}{c} d^2k \frac{1}{\lambda \sin\lambda d}$$
$$\times\left\{\left(\frac{\omega^2}{c^2}+k^2\right)\sin\lambda z \sin\lambda(z-d) + \lambda^2 \cos\lambda z \cos\lambda(z-d)\right\} \qquad \text{(A-7).}$$

With the application of complex frequency rotation ($\omega \to i\omega$),

$$\left\langle T^{00}\right\rangle_{in} = -\frac{\hbar c}{2(2\pi)^3}\int \frac{d\omega}{c} d^2k \frac{1}{\lambda \sinh\lambda d}$$
$$\times\left\{\left(-\frac{\omega^2}{c^2}+k^2\right)\sinh\lambda z \sinh\lambda(z-d) + \lambda^2 \cosh\lambda z \cosh\lambda(z-d)\right\} \qquad \text{(A-8).}$$

After appropriate change of variables and simple integral calculation, the inside energy per unit area is found as:

$$\frac{E_{in}}{area} = \frac{1}{area}\int \left\langle \hat{T}^{00}\right\rangle_{in} d^3x = \int_0^d \left\langle \hat{T}^{00}\right\rangle_{in} dz = -\frac{\hbar c}{6(2\pi)^2}\int_0^{\frac{\alpha}{2d}} \lambda^2\big((\lambda d)\coth\lambda d + 5\big)d\lambda =$$
$$= -\frac{\pi^2 \hbar c}{1440 d^3} I(\alpha) \qquad \text{(A-9),}$$

in which

$$I(\alpha) = \frac{3.75}{\pi^4}\int_0^{\alpha} x^2\left(x\frac{e^x+1}{e^x-1}+10\right)dx, \quad x = 2\lambda d \qquad \text{(A-10).}$$

As we know, even in the precise measurements (e.g. [10]), there is no direct experiment confirming the exact numerical coefficient in the already known result ($-\frac{\pi^2 \hbar c}{1440\, d^3}$) for the scalar field Casimir pressure because there are experimental difficulties in making two plates parallel at the scales and precisions needed in the modern experiments and unavoidable errors due to working with good real materials than perfectly ideal conductors. By the way, one can recover the ideal result by putting $I(\alpha) = 1$ which can be achieved by choosing $\alpha \approx 1.874786$ in (A-9); this is an acceptable value for $\alpha$ based on what explained following the relation (1).